\documentclass{desyproc}

\begin{document}
%------------------------------------
\title{%
  \vspace{-4\baselineskip}\hfill\textbf{\small DESY 13-178}\\[3\baselineskip]
Overview on Low-flux Detectors}

%for single authors the superscripts are optional
\author{{\slshape Jan Eike von Seggern}\\[1ex]
Deutsches Elektronen-Synchrotron (DESY), Hamburg, Germany}

% if the proceedings are available online (e.g. at Indico)
% please enter the contribution ID or file_name below for the DOI
%\contribID{32}
\contribID{seggern\_jan\_eike\_von}

% TO THE CONFERENCE EDITORS: 
% please update the following information      
% before sending the template to the authors
% \confID{800}  % if the conference is on Indico uncomment this line
\desyproc{DESY-PROC-2013-XX}
\acronym{Patras 2013} % if you want the Acronym in the page footer uncomment this line
% \doi  % if there is an online version we will register DOIs

\maketitle

\begin{abstract}
  Laboratory based searches for weakly-interacting slim particles
  (WISPs) of the light-shining-through-a-wall type (LSW) use visible or
  near-infrared (NIR) laser light. Low-noise and highly efficient
  detectors are necessary to improve over previous experiments. These
  requirements overlap with the requirements for single-photon detectors
  (SPDs) for quantum information (QI) experiments. In this contribution,
  the sensitivity of several QI SPDs is compared to photo-multiplier
  tubes (PMTs) and imaging charge-coupled devices (CCDs). It is found
  that only transition edge sensors (TESs) are viable alternatives to
  CCDs if the signal can be focussed to a few $\mathrm{\mu m}$.
\end{abstract}

\section{Introduction}

LSW experiments search for WISPs via the process $\gamma\to
\mathrm{wisp} \to\gamma$~\cite{Redondo:2010dp}. For a photon-counting
detection scheme, the signal rate, $\dot N_\mathrm{sig}$, is given by
\begin{equation*}
  \dot N_\mathrm{sig} =
  \dot N_\mathrm{in} \,
  \mathcal{P}(\gamma\to\mathrm{wisp}) \, 
  \mathcal{P}(\mathrm{wisp}\to\gamma) \, 
  \eta,
\end{equation*}
where $\dot N_\mathrm{in}$ is the rate of photons fed into the
experiment, 
$\mathcal{P}(\gamma\to\mathrm{wisp})$, $\eta$ the efficiency of the
detector and
$\mathcal{P}(\mathrm{wisp}\to\gamma)$ the probability for photon-WISP
and WISP-photon conversion, respectively, which are both proportional to
the square of the photon-WISP coupling, $g$. Hence, 
the sensitivity on
the coupling, $S(g)$, i.e.~the expected upper limit on $g$ for the case
that $g=0$ is realized in Nature, scales with the detector parameters
as
\begin{equation*}
  S(g) \propto ({\dot N_\mathrm{ul} }/{ \eta })^{1/4},
\end{equation*}
with $\dot N_\mathrm{ul}$ the count-rate sensitivity and $\eta$ the
quantum efficiency of the detector. The count rate sensitivity is
typically roughly proportional to the square root of the dark count
rate, $\dot N_\mathrm{ul} \propto \sqrt{\dot N_\mathrm{dc}}$. Hence, the
sensitivity can be improved (i.e.~lowered) by decreasing the dark count
rate or increasing the quantum efficiency.

To compare different detectors, the figure of merit $\mu = \eta/\dot
N_\mathrm{ul}$, is used. Thus, larger values of $\mu$ identify better
detectors. The count rate sensitivity is taken to be the average upper
limit of unified confidence intervals and is estimated using toy Monte
Carlo simulations~\cite{Feldman:1997qc}.

Early LSW experiments used PMTs for photo-detection~\cite{Ruoso:1992nx}.
Recent LSW experiments used CCDs and lasers in the visible
spectrum~\cite{Ehret:2010mh,Pugnat:2013dha}. Future LSW experiments will
use NIR lasers~\cite{ALPS:II} because optical elements are known to
withstand high powers at these wavelengths. At NIR wavelengths, silicon
based CCDs have a much reduced quantum efficiency compared to the
visible spectrum. Therefore, other devices for photo-detection are
sought. These detectors should have a quantum efficiency that is similar
to the quantum efficiency of CCDs in the optical and a dark count rate
below that of CCDs. Additionally, it is desirable that these detectors
can time-resolve single photons (SPD). A review of SPDs is given in
Ref.~\cite{eisamann:071101}.

In Sections~\ref{seggern:sec:ccd} to \ref{seggern:sec:pmt}
(electron-multiplying) CCDs, QI SPDs from Ref.~\cite{eisamann:071101}
and PMTs are discussed, respectively, and the figures of merit are
calculated. The results are compared in Section~\ref{seggern:sec:concl}.
To calculate the count rate sensitivity, $\dot N_\mathrm{ul}$, a
confidence level of $95\,\%$ is assumed.

\section{Imaging Charge-coupled Devices}
\label{seggern:sec:ccd}

CCDs are currently the prime choice for scientific visual imaging with a
wide range of devices to choose from. The imaging area of CCDs is
segmented into columns each consisting of a series of MIS\footnote{%
  metal-insulator-semiconductor structure
}
capacitors.
During data taking, these capacitors are biased into deep-depletion.
Incident photons are absorbed in the semiconductor material and produce
free charges which are stored by the capacitors. These charges are
integrated during an exposure. At the end of an exposure, the collected
charges are transported to a read-out structure and digitized. Hence, a
CCD cannot resolve single photons. In addition to the charges generated
by incident photons, thermally generated free charges are produced and
stored as well. These constitute the dark counts, which contributes
to the overall noise. The process of
read-out and digitization adds a second source of noise. Hence, the
total noise is given by
\begin{equation*}
  \sigma^2_\mathrm{tot} = \sigma^2_\mathrm{ro} + t \, R_\mathrm{dc},
\end{equation*}
where $\sigma_\mathrm{ro}$ is the read-out noise, $t$ the exposure time
and $R_\mathrm{dc}$ the production rate of dark counts.

% \hfill\strut%
%
\begin{wrapfigure}{O}{0.37\textwidth}
  \centering
  \includegraphics[width=0.37\textwidth]{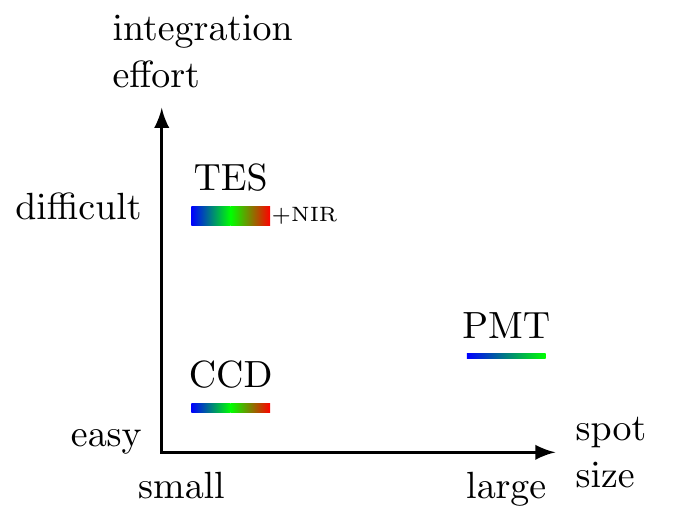}
  \caption{%
    Schematic summary of findings.
    The discussed detectors are ordered by integration effort and
    possible spot sizes. The color scales indicate the spectral range.
    The colorbars' relative sizes indicate the quantum efficiency.
  }
  \label{seggern:fig:findings}
\end{wrapfigure}%
The LSW experiment ALPS at DESY used a commercially available, low noise
CCD camera with $13\times13\,\mathrm{\mu m^2}$ sized
pixels (PIXIS CCD)~\cite{Ehret:2010mh,PIXIS:Datasheet}. A dark count rate below
$8\times10^{-4}\,\mathrm{e/(px\, s)}$ was achieved by liquid cooling of
the CCD chip and the camera was equipped with low-noise read-out
electronics ($\sigma_\mathrm{ro} =
4.3\,\mathrm{e}$)~\cite{Seggern:PhDThesis}. Thus, the read-out
noise is the larger contribution to the total noise for exposures
shorter than $1.5\,\mathrm{h}$. If the signal can be focussed to a
single pixel, a data set of 20 one hour exposures yields a figure of
merit $\mu = 1667 \,\mathrm{s/photon}$ for a quantum efficiency of
$80\,\%$ which is typical in the visible spectrum.

Electron multiplying CCDs (EMCCDs) amplify the charge signal before
read-out in an avalanche multiplication register~\cite{Madan:1983}. This
allows to neglect the read-out noise and, hence, short exposure times
are possible. But at the same time, the quantum efficiency is
effectively reduced by a factor of two due to the additional noise from
the multiplication process~\cite{Hadwen:2002}. The original quantum
efficiency, i.e.~without charge multiplication in an avalanche register,
can be recovered by interpreting the read-out values in a binary
fashion, i.e.~photon detected yes/no~\cite{Daigle:2009}, where a photon
is counted if the digitized signal is above a threshold, $k\,\sigma$.
Hence, the analysis can be reduced to that of a counting experiment. If
contamination by noise and loss of signal due to the threshold can be
neglected, this yields $\mu = 3307\,\mathrm{s/photon}$ assuming the same
values as above ($\eta=80\,\%$,
$R_\mathrm{dc}=8\times10^{-4}\,\mathrm{e/(px\, s)}$ and $20\,\mathrm{h}$
of data).

For NIR wavelengths, the PIXIS CCD was found to have a reduced quantum
efficiency of $1.2\,\%$~\cite{Seggern:PhDThesis}. The figure of merit is reduced accordingly for
the PIXIS CCD ($\mu = 24\,\mathrm{s/photon}$) and EMCCD ($\mu =
50\,\mathrm{s/photon}$). InGaAs based CCDs exist, which have a much
smaller band gap than silicon and, therefore, a much higher quantum
efficiency ($\sim85\,\%$) than the silicon-based PIXIS CCD. But these
devices also have a dark count rate, which is six orders of magnitude
above that of the PIXIS CCD~\cite{PI:InGaAs}. Therefore, these
specialized CCDs are of no help when improving the detector part of LSW
experiments. 

\section{Quantum Information Photo-detectors}

To maintain a low dark count rate and achieve a high quantum efficiency
at NIR wavelengths at the same time, sensors operated at cryogenic
temperatures can be used. Most of the devices listed in
Ref.~\cite{eisamann:071101} (cryogenic or not) have however dark count
rates much above that of the PIXIS CCD. 
Only 
transition edge sensors (TES)
were found to have low dark count rates below
that of the PIXIS CCD~\cite{TES}.
TES are bolometric sensors which are operated at
$\mathcal{O}(50\,\mathrm{mK})$.
Combined with a proper coating, high
quantum efficiencies of $95\,\%$ can be reached~\cite{Lita:2008}. The
dark count rate and quantum efficiency expected for ALPS-II
($\eta=75\,\%$ $\dot N_\mathrm{dc}=10^{-5}\,\mathrm{s}^{-1}$) are
assumed here as benchmark parameters~\cite{ALPS:II}. The corresponding figure of merit
for 20 hours of data is $\mu=14045\,\mathrm{s/photon}$.

\section{Photo-multiplier Tubes}
\label{seggern:sec:pmt}

The sensitive area of TES detectors and the pixels of a CCD are both of
order $\mathcal{O}(10\times10\,\mathrm{\mu m}^2)$. If the signal cannot
be focussed on such a small area, the pixels of a CCD can be binned.
But, as discussed above, the integrated dark count rate increases at the
same rate as the area of interest. Accordingly, the figure of merit and
the sensitivity on the coupling may worsen significantly. In this case,
PMTs are a very good alternative although they have a limited quantum
efficiency ($\eta \lesssim 30\,\%$) and a limited spectral range
($300\,\mathrm{nm}\le \lambda \le 850
\,\mathrm{nm}$)~\cite{PMT:Handbook}. The sensitive area of a
% \begin{table}
\begin{wraptable}{o}{0pt}
  \centering
  \begin{tabular}{|lccc|}
    \hline
    Detector & $\eta$ [\%] & $\dot N_\mathrm{dc}$ [$\mathrm{s}^{-1}$] &
    $\mu$ [$\mathrm{s/photon}$]
    \\
    \hline
    CCD (visible)
    & 80\phantom{.0} & $8\times 10^{-4}$ & \phantom{0}1667
    \\
    EMCCD (visible)
    & 80\phantom{.0} & $8\times 10^{-4}$ & \phantom{0}3307
    \\
    CCD (NIR)
    & \phantom{0}1.2 & $8\times 10^{-4}$ & \phantom{000}24
    \\
    EMCCD (NIR)
    & \phantom{0}1.2 & $8\times 10^{-4}$ & \phantom{000}51
    \\
    TES
    & 75\phantom{.0} & $10^{-5}$ & 14045 
    \\
    PMT
    & 25\phantom{.0} & 0.5 & \phantom{000}39
    \\
    \hline
  \end{tabular}
  \caption{%
    Comparison of different detectors.
    The table lists the typical quantum efficiency, $\eta$, and dark
    count rate, $\dot N_\mathrm{dc}$,
    together with the figure of merit, $\mu$, for silicon CCD/EMCCD, TES
    and PMT as discussed in the text.
  }
  \label{seggern:tab:comparison}
\end{wraptable}
% \end{table}
PMT consists of a photo-sensitive material with a low work function.
Incident photons produce free electrons which are directed to an
electron multiplier by a focussing electrode. The high gain of the
electron multiplier allows single photon detection. Cooling the
sensitive area reduces the dark count rate. For example, the SHIPS
helioscope uses a PMT with an active area of $2.5\,\mathrm{cm}^2$, which
has a peak quantum efficiency of $25\,\%$ and a dark count rate of
$0.5\,\mathrm{cnt/s}$ when cooled to
$-21^\circ\mathrm{C}$~\cite{Schwarz}. This corresponds to $\mu =
39\,\mathrm{s/photon}$.

\section{Conclusion}
\label{seggern:sec:concl}

Surprisingly, of all SPDs used in QI experiments, only TES detectors
have a sufficiently low dark count rate to improve significantly over
conventional CCDs. The figures of merit of the detectors mentioned in
the above sections are summarized in Tab.~\ref{seggern:tab:comparison}.
Of the presented alternatives, a TES is the best option. Especially in
the NIR, a TES is superior to a CCD because its quantum efficiency does
not deteriorate for these wavelengths. In the visible regime, CCDs
remain a viable option when only few resources are available for
detector development. From the values listed in
Tab.~\ref{seggern:tab:comparison}, it seems that PMTs are the worst
option. Their figure of merit is two orders of magnitude below that of
CCDs (visible), which is caused mainly by their high dark count rate.
However, considering their large sensitive area, PMTs are the detector
of choice if the signal cannot be focussed very well. These findings are
schematically summarized in Fig.~\ref{seggern:fig:findings}.

\section*{Acknowledgments}

The study of the PIXIS CCD was kindly supported by G.~Wiedemann
(Hamburger Sternwarte).

% ****************************************************************************
% BIBLIOGRAPHY AREA
% ****************************************************************************

\begin{footnotesize}

\end{footnotesize}

% ****************************************************************************
% END OF BIBLIOGRAPHY AREA
% ****************************************************************************

\end{document}